\documentclass{elsart}
\usepackage{graphicx}
\usepackage{amssymb}

\usepackage{amsmath}

\DeclareMathOperator{\tr}{tr}

\begin{document}

\begin{frontmatter}



\title{Restoration of supersymmetry on the lattice:\\
Two-dimensional $\mathcal{N}=(2,2)$ supersymmetric Yang-Mills theory}


\author{Issaku Kanamori},
\ead{kanamori-i@riken.jp}
\author{Hiroshi Suzuki}
\ead{hsuzuki@riken.jp}
\address{Theoretical Physics Laboratory, RIKEN, Wako 2-1, Saitama 351-0198,
Japan}

\begin{abstract}
% Text of abstract
By numerically investigating the conservation law of the supercurrent, we
confirm the restoration of supersymmetry in Sugino's lattice formulation of the
two-dimensional $\mathcal{N}=(2,2)$ supersymmetric $SU(2)$ Yang-Mills theory
with a scalar mass term. Subtlety in the case without the scalar mass term,
that appears to ruin perturbative power counting, is also pointed out.
\end{abstract}

\end{frontmatter}

% main text
%\section{}
%\label{}
\section{Introduction}
\label{sec:1}
Nonperturbative study of supersymmetric gauge theory is of great general
interest but in the context of lattice formulation, no compelling evidence of
supersymmetry has been observed so far. Spacetime lattice is generally
irreconcilable with supersymmetry and one must fine-tune coefficients of
relevant and marginal operators so that supersymmetric Ward-Takahashi (WT)
identities are restored in the continuum limit. (An important exception is the
four-dimensional $\mathcal{N}=1$ supersymmetric Yang-Mills theory
(SYM)~\cite{Curci:1986sm}; see Ref.~\cite{Montvay:2001aj} for a lot of effort
went into numerical study of this system; see Ref.~\cite{Giedt:2008cd} for a
recent attempt.) Recently, for two- and three-dimensional extended
supersymmetric gauge theories, lattice formulations that require no (or a
little) fine tuning have been
proposed~\cite{Kaplan:2002wv,Cohen:2003xe,Cohen:2003qw,Sugino:2003yb,%
Sugino:2004qd,Sugino:2004uv,Catterall:2004np,Kaplan:2005ta,D'Adda:2005zk,%
Sugino:2006uf,Endres:2006ic,D'Adda:2007ax,Nagata:2008zz,Matsuura:2008cf,%
Sugino:2008yp}.\footnote{For relations among these formulations, see
Refs.~\cite{Unsal:2006qp,Damgaard:2007be,Takimi:2007nn,Damgaard:2007xi,%
Damgaard:2007eh,Catterall:2007kn,Damgaard:2008pa}.} It can be argued that exact
fermionic symmetries in these lattice formulations, combined with other lattice
symmetries, (almost) prohibit relevant and marginal supersymmetry breaking
operators to appear~\cite{Cohen:2003xe,Sugino:2003yb}. Supersymmetry is then
restored in the continuum limit without (or with a little) fine
tuning.\footnote{See also Refs.~\cite{Suzuki:2005dx,Elliott:2005bd} for
alternative approaches that do not use exact fermionic symmetries.}

The aim of the present study is to test this scenario of supersymmetry
restoration, in Sugino's lattice formulation of the two-dimensional
$\mathcal{N}=(2,2)$ $SU(2)$ SYM~\cite{Sugino:2003yb,Sugino:2004qd}. By a Monte
Carlo simulation, we study a supersymmetric WT identity in the form of the
supercurrent conservation law. For the reason elucidated in Sec.~\ref{sec:4},
we introduce a supersymmetry breaking scalar mass term to the lattice
formulation. The expected WT identity hence takes the form of a partially
conserved supercurrent (``PCSC'') relation in which the breaking term is
proportional to the square of the scalar mass. We numerically confirm the
restoration of this PCSC relation in the continuum limit. Our result strongly
indicates that the proposed scenario for the supersymmetry restoration is in
fact valid and the target continuum theory (i.e., the two-dimensional
$\mathcal{N}=(2,2)$ $SU(2)$ SYM with a soft supersymmetry breaking mass term)
is realized in the continuum limit of the present lattice model.

\section{Preliminaries in the continuum theory}
\label{sec:2}
The euclidean continuum action of the two-dimensional $\mathcal{N}=(2,2)$
$SU(N_c)$ SYM, that is obtained by dimensionally reducing the $\mathcal{N}=1$
SYM from four to two dimensions, is given by\footnote{In this paper, we adopt
the representation
\begin{equation}
   \Gamma_0=\begin{pmatrix}
     -i\sigma_1&0\\
     0&i\sigma_1
   \end{pmatrix},\quad
   \Gamma_1=\begin{pmatrix}
     i\sigma_3&0\\
     0&-i\sigma_3
   \end{pmatrix},\quad
   \Gamma_2=\begin{pmatrix}
     0&-i\\
     -i&0
   \end{pmatrix},\quad
   \Gamma_3=C=\begin{pmatrix}
     0&1\\
     -1&0
   \end{pmatrix},
\end{equation}
and $\Gamma_5\equiv\Gamma_0\Gamma_1\Gamma_2\Gamma_3=
\begin{pmatrix}
     \sigma_2&0\\
     0&-\sigma_2
   \end{pmatrix}$, and set
\begin{equation}
   \Psi^T=(\psi_0,\psi_1,\chi,\eta/2)
\label{two}
\end{equation}
that corresponds to the conventional basis in the topological field
theory~\cite{Witten:1988ze,Witten:1990bs}. See also~Ref.~\cite{Kato:2003ss}.}
\begin{equation}
   S=\frac{1}{g^2}
   \int d^2x\,\tr\left\{
   \frac{1}{2}F_{MN}F_{MN}+\Psi^TC\Gamma_MD_M\Psi+\widetilde H^2\right\},
\label{three}
\end{equation}
where Roman indices~$M$ and $N$ run over 0, 1, 2 and 3, while Greek
indices~$\mu$ and~$\nu$ below run over only 0 and~1. $F_{MN}$ are field
strengths $F_{MN}=\partial_MA_N-\partial_NA_M+i[A_M,A_N]$ and covariant
derivatives $D_M$ are defined with respect to the adjoint representation,
$D_M\Psi=\partial_M\Psi+i[A_M,\Psi]$. Here, in all expressions, it is
understood that $\partial_2\to0$ and~$\partial_3\to0$ (dimensional reduction).
We also define complex scalar fields, $\phi\equiv A_2+iA_3$
and~$\overline\phi\equiv A_2-iA_3$. The auxiliary field $\widetilde H$ is
related to the auxiliary field~$H$ in Refs.~\cite{Sugino:2003yb,Sugino:2004qd}
by $\widetilde H\equiv H-iF_{01}$. We also introduce a soft supersymmetry
breaking term
\begin{equation}
   S_{\text{mass}}=\frac{1}{g^2}\int d^2x\,\mu^2
   \tr\left\{\overline\phi\phi\right\}.
\label{four}
\end{equation}
This term is ``soft'' in the sense that it does not introduce new ultraviolet
divergences compared with the supersymmetric theory~$S$.

Action~(\ref{three}) is invariant under the super transformation,
$\delta A_M=i\epsilon^TC\Gamma_M\Psi$,
$\delta\Psi=\frac{i}{2}F_{MN}\Gamma_M\Gamma_N\epsilon
+i\widetilde H\Gamma_5\epsilon$
and $\delta\widetilde H=-i\epsilon^TC\Gamma_5\Gamma_MD_M\Psi$, where the
parameter~$\epsilon$ has four spinor components, $\epsilon^T\equiv
-(\varepsilon^{(0)},\varepsilon^{(1)},\widetilde\varepsilon,\varepsilon)$.
We define components of the super transformation by
\begin{equation}
   \delta\equiv\varepsilon^{(0)}Q^{(0)}+\varepsilon^{(1)}Q^{(1)}
   +\widetilde\varepsilon\widetilde Q+\varepsilon Q.
\label{five}
\end{equation}
Also, the total action~$S+S_{\text{mass}}$ possesses following global symmetries:
The $U(1)_A$ symmetry,
\begin{equation}
   \Psi\to\exp\left\{\alpha\Gamma_2\Gamma_3\right\}\Psi,\quad
   \phi\to\exp\left\{2i\alpha\right\}\phi,\quad
   \overline\phi\to\exp\left\{-2i\alpha\right\}\overline\phi,
\label{six}
\end{equation}
the $U(1)_V$ symmetry,
\begin{equation}
   \Psi\to\exp\left\{i\alpha\Gamma_5\right\}\Psi,
\label{seven}
\end{equation}
and a global $Z_2$ symmetry,
\begin{equation}
   \Psi\to i\Gamma_2\Psi,\quad
   \phi\to-\overline\phi,\quad
   \overline\phi\to-\phi.
\label{eight}
\end{equation}
The last one is a remnant of the reflection symmetry with respect to the $M=2$
direction of the four-dimensional theory before dimensional reduction.

A Noether current associated with the supersymmetry of $S$ (the supercurrent)
is given by
\begin{equation}
   s_\mu\equiv
   -\frac{1}{g^2}C\Gamma_M\Gamma_N\Gamma_\mu\tr\left\{F_{MN}\Psi\right\}
   \equiv\left(\mathcal{J}_\mu^{(0)},\mathcal{J}_\mu^{(1)},
   \widetilde{\mathcal{J}}_\mu,\mathcal{J}_\mu\right)^T.
\label{nine}
\end{equation}
These four spinor components correspond to fermionic transformations
in~Eq.~(\ref{five}), $Q^{(0)}$, $Q^{(1)}$, $\widetilde Q$ and~$Q$, respectively.
If the auxiliary field~$H$, instead of $\widetilde H$, is regarded as an
independent dynamical variable as is the case below, $F_{01}=-F_{10}$ in this
expression must be understood as~$-iH$.

We will consider a correlation function of the supercurrent and another
fermionic operator. As a lowest-dimensional gauge invariant fermionic operator,
we take
\begin{equation}
   f_\mu\equiv
   \frac{1}{g^2}\Gamma_\mu\left(
   \Gamma_2\tr\{A_2\Psi\}+\Gamma_3\tr\{A_3\Psi\}\right)
   \equiv\left(X_\mu^{(0)},X_\mu^{(1)},\widetilde{X}_\mu,X_\mu\right)^T.
\label{ten}
\end{equation}
This is gauge invariant because the scalar fields $A_2$ and $A_3$, as well as
the spinor~$\Psi$, transform as the adjoint representation under
two-dimensional gauge transformations.
We in particular consider a ``diagonal part'' in the product of
supercurrent~(\ref{nine}) and operator~(\ref{ten}). That is, we consider
following four correlation functions
\begin{equation}
   \left\langle\mathcal{J}_\mu^{(0)}(x)X_\nu^{(0)}(0)\right\rangle,\quad
   \left\langle\mathcal{J}_\mu^{(1)}(x)X_\nu^{(1)}(0)\right\rangle,\quad
   \left\langle\widetilde{\mathcal{J}}_\mu(x)\widetilde X_\nu(0)\right\rangle,
   \quad
   \left\langle\mathcal{J}_\mu(x)X_\nu(0)\right\rangle.
\label{eleven}
\end{equation}
It is useful to note that these four correlation functions are interchanged
each other under transformations~(\ref{seven}) with $\alpha=\pi/2$
and~(\ref{eight}). Since Eqs.~(\ref{seven}) and~(\ref{eight}) are symmetries of
the present system, four correlation functions in~Eq.~(\ref{eleven}) coincide
in the continuum theory. On the other hand, the present lattice formulation is
not invariant under either Eq.~(\ref{seven}) or Eq.~(\ref{eight}) (while
Eq.~(\ref{six}) is manifestly realized) and a ``degeneracy'' of these four
correlation functions can be used as a useful measure for how we are close to
the continuum.

With the presence of the supersymmetry breaking
term~$S_{\text{mass}}$~(\ref{four}), the supercurrent is not conserved and,
defining
\begin{equation}
   f\equiv
   -2C
   \left(\Gamma_2\tr\{A_2\Psi\}+\Gamma_3\tr\{A_3\Psi\}\right)
   \equiv\left(Y^{(0)},Y^{(1)},\widetilde{Y},Y\right)^T,
\label{twelve}
\end{equation}
we obtain the supersymmetric WT identity (the PCSC relation) in spinor
components
\begin{gather}
   \partial_\mu\left\langle\mathcal{J}_\mu^{(0)}(x)X_\nu^{(0)}(0)\right\rangle
   -\frac{\mu^2}{g^2}\left\langle Y^{(0)}(x)X_\nu^{(0)}(0)\right\rangle
   =-i\delta^2(x)\left\langle Q^{(0)}X_\nu^{(0)}(0)\right\rangle,
\label{thirteen}\\
   \partial_\mu\left\langle\mathcal{J}_\mu^{(1)}(x)X_\nu^{(1)}(0)\right\rangle
   -\frac{\mu^2}{g^2}\left\langle Y^{(1)}(x)X_\nu^{(1)}(0)\right\rangle
   =-i\delta^2(x)\left\langle Q^{(1)}X_\nu^{(1)}(0)\right\rangle,
\label{fourteen}\\
   \partial_\mu\left\langle
   \widetilde{\mathcal{J}}_\mu(x)\widetilde X_\nu(0)\right\rangle
   -\frac{\mu^2}{g^2}\left\langle
   \widetilde{Y}(x)\widetilde X_\nu(0)\right\rangle
   =-i\delta^2(x)\left\langle\widetilde Q\widetilde X_\nu(0)\right\rangle,
\label{fifteen}\\
   \partial_\mu\left\langle\mathcal{J}_\mu(x)X_\nu(0)\right\rangle
   -\frac{\mu^2}{g^2}\left\langle Y(x)X_\nu(0)\right\rangle
   =-i\delta^2(x)\left\langle QX_\nu(0)\right\rangle,
\label{sixteen}
\end{gather}
if the regularization preserves supersymmetry. We emphasize that these local
relations hold irrespective of boundary conditions. One can derive these
relations in the functional integral by employing a \emph{local\/} change of
variables that does not ``touch'' the boundary. See Appendix~\ref{appendix:A}.
Therefore, these hold in particular with the antiperiodic temporal boundary
condition for fermionic variables, although this boundary condition explicitly
breaks supersymmetry (for example, the energy spectrum would not be
supersymmetric with this boundary condition).

We now argue that, if the argument for the supersymmetry restoration
in~Ref.~\cite{Sugino:2003yb} is valid,
relations~(\ref{thirteen})--(\ref{sixteen}) must be reproduced for $x\neq0$ in
the continuum limit of the lattice model.

According to the argument of Ref.~\cite{Sugino:2003yb}, the lattice action in
Ref.~\cite{Sugino:2003yb} provides a regularization, that becomes
supersymmetric in the continuum limit, for all 1PI correlation function of
\emph{elementary fields}. In particular, possible ultraviolet divergent
functions, tadpoles and the scalar self-energy at the one-loop level, take the
form (in the continuum limit) expected in the continuum theory (i.e., tadpoles
vanish and no self-energy correction at zero external momentum). In this sense,
the present lattice regularization is supersymmetric. This almost implies that
supersymmetric WT identities such as Eqs.~(\ref{thirteen})--(\ref{sixteen})
hold in the continuum limit.

Eqs.~(\ref{thirteen})--(\ref{sixteen}), however, contain \emph{composite
operators\/} and a definition of composite operators in a lattice formulation
is to a large extent arbitrary. We thus must be sure that lattice artifacts,
when combined with ultraviolet divergences arising from these operators, do not
modify the WT identities. Fortunately, in the present two-dimensional
super-renormalizable system, operators in Eqs.~(\ref{nine}), (\ref{ten})
and~(\ref{twelve}) themselves are ultraviolet finite (i.e., no operator
renormalization/mixing is required; this significantly simplifies consideration
of the supersymmetry WT identity compared with four dimensions).\footnote{Only
possible operator that the supercurrent can mix with (in the one-loop level)
is $C\Gamma_\mu\tr\{\Psi\}$ and this identically vanishes for the gauge
group~$SU(N_c)$.} Only possible ultraviolet divergence in~Eq.~(\ref{sixteen}),
for example, arises from a one-loop diagram obtained by connecting elementary
fields in~$\mathcal{J}_\mu$ and~$X_\nu$. However, this ultraviolet divergence
can readily be avoided by setting two points $x$ and~$0$ to be
separated.\footnote{This is impossible for one-point WT identities studied in
Refs.~\cite{Catterall:2006jw,Catterall:2006is,Suzuki:2007jt,Fukaya:2007ci};
these one-point functions are thus subject of lattice artifacts and cannot
directly be used to observe the supersymmetry restoration.}
In fact, by analysing this one-loop contribution, one has
\begin{align}
   &\partial_\mu\left\langle\mathcal{J}_\mu(x)X_\nu(0)\right\rangle
   -\frac{\mu^2}{g^2}\left\langle Y(x)X_\nu(0)\right\rangle
\nonumber\\
   &=-i\delta^2(x)\left\langle QX_\nu(0)\right\rangle
   +\frac{1}{4\pi}(N_c^2-1)(c_Q-1)\partial_\nu\delta^2(x)
\label{seventeen}
\end{align}
for Eq.~(\ref{sixteen}) and similar corrections for
Eqs.~(\ref{thirteen})--(\ref{fifteen})
(the constant~$c_Q$ here can differ for each of relations, as $c_{Q^{(0)}}$,
$c_{Q^{(1)}}$ and $c_{\widetilde Q}$, depending on the regularization),
assuming that the regularization for 1PI functions of elementary fields
preserves supersymmetry. The constant~$c_Q$ in~Eq.~(\ref{seventeen}) depends on
the definition (or regularization) of composite operators $\mathcal{J}_\mu$ and
$X_\nu$, but this dependence on regularization disappears
for~$x\neq0$.\footnote{Incidentally, the last term of Eq.~(\ref{seventeen}) is
not a genuine anomaly of supersymmetry. In fact, there exists a possible
definition of $\mathcal{J}_\mu$ that leads to $c_Q=1$; see
footnote~\ref{footnote:12}.}

In summary, if the argument for the supersymmetry restoration in our lattice
model is valid, relations~(\ref{thirteen})--(\ref{sixteen}) with~$x\neq0$ must
be reproduced in the continuum limit.\footnote{Our argumentation here is not
quite rigorous. For more satisfactory treatment, one first derives a WT
identity associated with an appropriate (would-be) supersymmetry transformation
in lattice theory and then shows all lattice artifacts to
Eqs.~(\ref{thirteen})--(\ref{sixteen}) with $x\neq0$ vanish in the continuum
limit.}

One might wonder why we worry about the supersymmetry restoration in the
present model; after all, supersymmetry is explicitly broken by the scalar mass
term (and by the boundary condition that we will adopt below). It is very
important, however, to distinguish three different possible sources for
supersymmetry breaking in the present lattice model; the scalar mass term, the
boundary condition and the lattice regularization itself. Our point is that the
observation of WT identities~(\ref{thirteen})--(\ref{sixteen}) enables us to
\emph{isolate\/} the last source of supersymmetry breaking. What we are talking
about is whether this supersymmetry breaking owing to the lattice
regularization disappears in the continuum limit or not.

\section{Results of Monte Carlo study}
\label{sec:3}
For details of the lattice formulation in
Refs.~\cite{Sugino:2003yb,Sugino:2004qd}, we refer the reader to the original
references and Refs.~\cite{Suzuki:2007jt,Kanamori:2007yx}. The point is that
the fermionic transformation $Q$ in Eq.~(\ref{five}) is nilpotent (up to gauge
transformations) and action~(\ref{three}) can be written as a $Q$-exact form,
as topological field theory~\cite{Witten:1988ze,Witten:1990bs}. These
nilpotent~$Q$ and $Q$-exact action, and thus the invariance of the action
under~$Q$, can be realized even in lattice gauge
theory~\cite{Sugino:2003yb,Sugino:2004qd}.
This sort of exact fermionic symmetry is realized also in lattice formulations
of the present system in~Refs.~\cite{Kaplan:2002wv,Cohen:2003xe,%
Catterall:2004np,D'Adda:2005zk,Sugino:2006uf}.
Other fermionic symmetries,
$Q^{(0)}$, $Q^{(1)}$ and~$\widetilde Q$, from which
Eqs.~(\ref{thirteen})--(\ref{fifteen}) follow, are not preserved in the lattice
formulation and the question is whether the invariance under these is restored
in the continuum limit.

We consider a finite two-dimensional rectangular lattice
\begin{equation}
   \Lambda\equiv\left\{x\in a\mathbb{Z}^2\mid 0\leq x_0<\beta\equiv2L,\,\,
   0\leq x_1<L\right\},
\end{equation}
where $a$ is the lattice spacing and the physical size $L$ is fixed to be
$\sqrt{2}/g$.\footnote{Note that in two dimensions, the gauge coupling~$g$ has
the mass dimension~1.} Except the result in Fig.~\ref{fig:7} below, all the
results were obtained with the antiperiodic temporal ($=x_0$) boundary
condition for fermionic variables (the $x_1$-direction is always periodic).
We found that this boundary condition leads to faster approach to the
continuum and less noisy signals, compared with the periodic boundary condition.
Recall that WT identities~(\ref{thirteen})--(\ref{sixteen}) in the continuum
theory must hold irrespective of the boundary condition.\footnote{In the
present lattice formulation, the invariance under the fermionic
transformation~$Q$ is manifest. Then, one can derive an exact WT identity of
the form~(\ref{sixteen}) in lattice theory with an appropriate definition of
composite operators (see footnote~\ref{footnote:12}). The
boundary~$x_0=0\equiv\beta$, to which the antiperiodic boundary condition
refers, can freely be shifted by change of fermionic variables. This also
illustrates that the boundary condition is irrelevant to local WT identities
such as~(\ref{thirteen})--(\ref{sixteen}).}

Corresponding to mass term~(\ref{four}), we add a scalar mass term
\begin{equation}
   \frac{\mu^2}{g^2}\sum_{x\in\Lambda}\tr\left\{\overline\phi(x)\phi(x)\right\}
\label{nineteen}
\end{equation}
to the lattice action of Ref.~\cite{Sugino:2004qd}.\footnote{We adopt a
convention that all lattice field variables are dimensionless; the continuum
dimensionful scalar field~$\phi$ is replaced by $\phi(x)/a$ on the lattice.}
For lattice transcription of operators in Eqs.~(\ref{nine}), (\ref{ten})
and~(\ref{twelve}), we adopt a simple prescription that a field variable at a
point~$x$ is replaced by the corresponding lattice field variable at~$x$.
Covariant derivatives for the scalar field are replaced by the forward
covariant differences,
$D_\mu\phi(x)\to\{U(x,\mu)\phi(x+a\hat\mu)U(x,\mu)^{-1}-\phi(x)\}/a^2$,
where $U(x,\mu)$ are link variables.
As noted in the preceding section, a precise way of lattice transcription of
these operators should be irrelevant to our present analysis. That is, all
ambiguities associated with the operator definition are integrated in the
constants~$c_Q$ in Eq.~(\ref{seventeen}) (and $c_{Q^{(0)}}$, $c_{Q^{(1)}}$
and~$c_{\widetilde Q}$) in the continuum limit and they do not appear
for~$x\neq0$.\footnote{\label{footnote:12}
By employing lattice perturbation theory and the
Reisz power counting theorem, one finds that the above definition of
$\mathcal{J}_\mu$ and~$X_\nu$ leads to $c_Q=1-\pi$ in~Eq.~(\ref{seventeen}).
Since the present lattice formulation has an exact fermionic symmetry~$Q$,
there exists an alternative natural definition of $\mathcal{J}_\mu$ that
exactly fulfills the corresponding WT identity even with finite lattice
spacings. This improved Noether current differs from the above definition by
terms of $O(a)$ and these terms improve the quantum property of the current. In
particular, in the continuum limit, one has $c_Q=1$ in~Eq.~(\ref{seventeen}).
Although this improved current has a desired property, here we do not adopt
this definition for several reasons. Firstly, our prime objective is to see the
restoration of symmetries other than~$Q$; there is no particularly superior
definition of Noether currents associated with $Q^{(0)}$, $Q^{(1)}$
and~$\widetilde Q$ in the present lattice formulation. Secondly, we want to
illustrate the idea that a precise form of the operators is irrelevant for
relations~(\ref{thirteen})--(\ref{sixteen}) with~$x\neq0$.}

One of us (I.~K.) developed a Rational Hybrid Monte Carlo
(RHMC)~\cite{Clark:2004cp} simulation code (on the basis of the
multi-shift~\cite{Frommer:1995ik} CG solver~\cite{Jegerlehner:1996pm}) for the
present lattice system with the gauge group $SU(2)$, by utilizing available
libraries and programs~\cite{DiPierro:2000bd,DiPierro:2005qx,Remez}. Some
details of this simulation code have already been reported
in~Ref.~\cite{Kanamori:2008vi}. (See also Fig.~3 of Ref.~\cite{Kanamori:2008ve}
that illustrates how the effect of dynamical fermions is properly included with
this code.)

We summarize on parameters in the present simulation: The parameter~$\epsilon$
for the admissibility~\cite{Sugino:2004qd} was fixed to
be~$2.6$.\footnote{Although a precise choice of the
parameter~$0<\epsilon<2\sqrt{2}$ should be irrelevant for results with small
lattice spacings, we admit that we have not carried out systematic study on
this point.
As a simple consistency check, we performed a small experiment that measures
the expectation value of the action density of the pure Yang-Mills part
(Eq.~(4.27) of~Ref.~\cite{Kanamori:2007yx}) for different values of $\epsilon$,
$\epsilon=1.0$, $2.0$, and~$2.6$. The scalar mass squared is~$\mu^2/g^2=0.25$,
lattice size is~$6\times6$, and the lattice spacing is $ag=0.2357$. The number
of uncorrelated configurations is 100 for each~$\epsilon$. We observed that the
admissibility bound $\|1-U(x,0,1)\|<\epsilon$ (see Ref.~\cite{Kanamori:2007yx})
was never exceeded in the molecular dynamics and the expectation values for
each $\epsilon$ coincided within statistical errors.}
The degree of the rational approximation in RHMC was typically~$20$--$30$. The
multi time step acceleration~\cite{Sexton:1992nu} was used in the molecular
dynamics (see~Appendix~\ref{appendix:B}).
The time interval of one trajectory was fixed to be~$0.5$ and time steps of
leapfrog were chosen such that the acceptance was greater than~$0.8$. We stored
configurations in every $10$~trajectories. The autocorrelation was then
estimated by jackknife analysis with binning. We discarded first $10000$
configurations for thermalization and used subsequent configurations at every
$50$ configurations. In this way, we prepared uncorrelated configurations
listed in Table~\ref{table:1}.
\begin{table}
\caption{Set of configurations we used.}
\label{table:1}
\begin{center}
\begin{tabular}{ccccc}
\hline\hline
 $\mu^2/g^2$ & lattice size & $ag$ & number of configurations & set label\\
\hline
0.04 & $12\times6$ & 0.2357 & 800 & I (a) \\
0.04 & $16\times8$ & 0.1768 & 800 & I (b) \\
0.04 & $20\times10$ & 0.1414 & 800 & I (c) \\
\hline
0.25 & $12\times6$ & 0.2357 & 800 & II (a) \\
0.25 & $16\times8$ & 0.1768 & 800 & II (b) \\
0.25 & $20\times10$ & 0.1414 & 800 & II (c) \\
\hline
0.49 & $12\times6$ & 0.2357 & 800 & III (a) \\
0.49 & $16\times8$ & 0.1768 & 1800 & III (b) \\
0.49 & $20\times10$ & 0.1414 & 1800 & III (c) \\
\hline
1.0 & $12\times6$ & 0.2357 & 800 & IV (a) \\
1.0 & $16\times8$ & 0.1768 & 1800 & IV (b) \\
1.0 & $20\times10$ & 0.1414 & 1800 & IV (c) \\
\hline
1.69 & $12\times6$ & 0.2357 & 800 & V (a) \\
1.69 & $16\times8$ & 0.1768 & 1800 & V (b) \\
1.69 & $20\times10$ & 0.1414 & 1800 & V (c) \\
\hline\hline
\end{tabular}
\end{center}
\end{table}

In the present lattice formulation, the pfaffian of the Dirac operator
resulting from the integration over fermionic variables, that is real positive
in the continuum theory, is generally complex owing to lattice artifacts.
For parameters in the present simulation with the antiperiodic boundary
condition, we found that the absolute value of the complex phase of the Dirac
determinant is $\lesssim0.8$ radians. We took into account the complex phase of
pfaffian by reweighting the half of the phase of the determinant. See
Refs.~\cite{Suzuki:2007jt,Kanamori:2007yx}. We observed that, however, this
reweighting had practically no notable effect. Also, all quantities presented
below, that are real in the continuum theory, are generally complex with finite
lattice spacings. The imaginary part is however generally small (and
consistent with zero) and we will present only the real part for simplicity.

For the correlation functions used in the following analyses, we took an
average over the whole lattice points to increase the effective number of
configurations. For example,
$\langle\mathcal{J}_0^{(0)}(x)X_0^{(0)}(0)\rangle$ is computed by taking the
average of $\langle\mathcal{J}_0^{(0)}(y)X_0^{(0)}(z)\rangle$ over all
$y\in\Lambda$ and~$z\in\Lambda$ with
$y_0-z_0=x_0\bmod\beta$ and~$y_1-z_1=x_1\bmod L$ are kept fixed.\footnote{
With the understanding that the sign of the correlation function is flipped
when $y-z=x-\beta\hat0$, the correlation function is translationally invariant
even with the antiperiodic boundary condition.}

\begin{figure}[htbp]
\begin{center}
\includegraphics*[width=\textwidth,clip]{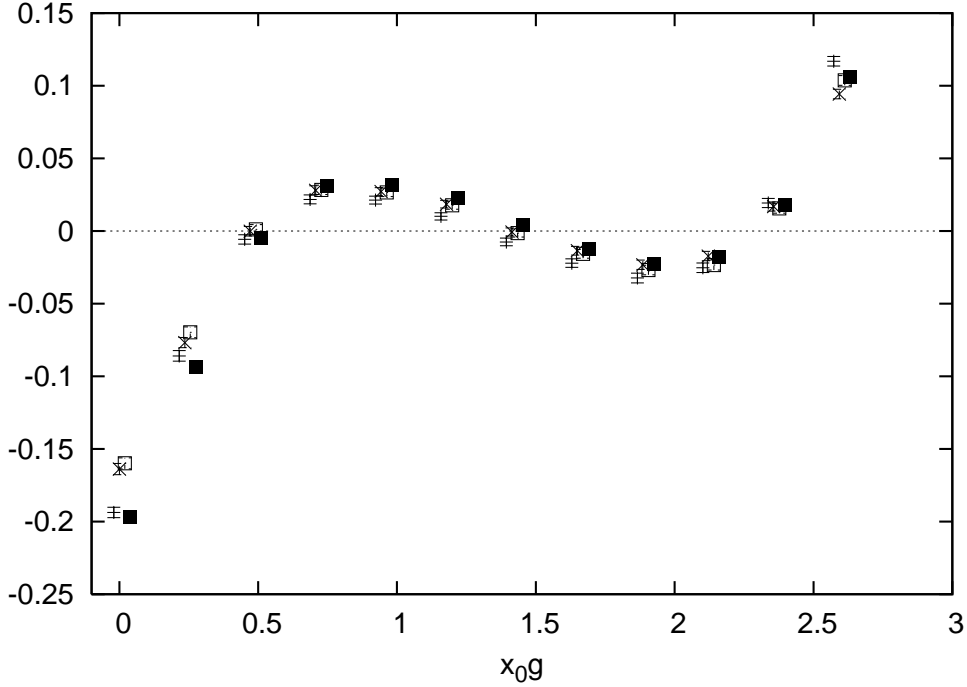}
\end{center}
\caption{Correlation functions~(\ref{eleven}) with $\mu=\nu=0$ along the
line~$x_1=L/2$ obtained by set~IV~(a) of Table~\ref{table:1}:
$\langle\mathcal{J}_0^{(0)}(x)X_0^{(0)}(0)\rangle/g^2$~($+$),
$\langle\mathcal{J}_0^{(1)}(x)X_0^{(1)}(0)\rangle/g^2$~($\times$),
$\langle\widetilde{\mathcal{J}}_0(x)\widetilde X_0(0)\rangle/g^2$~($\square$),
$\langle\mathcal{J}_0(x)X_0(0)\rangle/g^2$~($\blacksquare$).}
\label{fig:1}
\end{figure}

As a typical example of correlation functions, we plot in Fig.~\ref{fig:1}
correlation functions~(\ref{eleven}) obtained by employing set~IV~(a)
of~Table~\ref{table:1} as a function of~$x_0$ ($x_1=L/2$). Four correlation
functions are quite well-degenerated, except at points near $x=0$ and its
periodic images. (As noted, at $x=0$ and its periodic images, the correlation
functions suffer from lattice artifacts associated with the operator
definition.) We thus have an indication that the lattice spacing~$ag=0.2357$ is
already rather close to the continuum, at least for the correlation functions
with $\mu^2/g^2=1.0$.

\begin{figure}[htbp]
\begin{center}
\includegraphics*[width=\textwidth,clip]{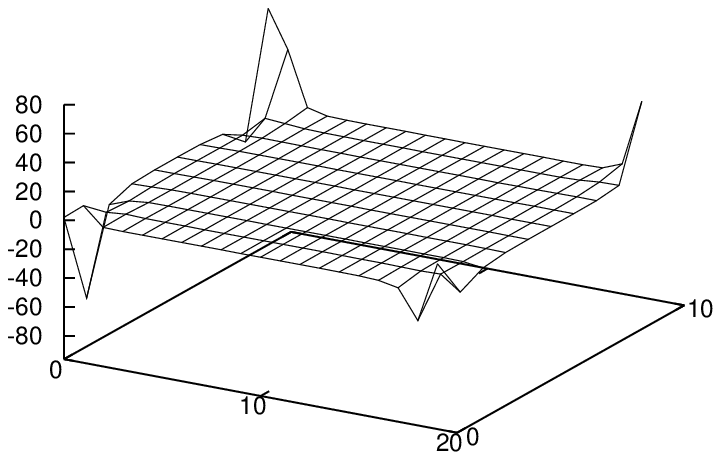}
\end{center}
\caption{A bird's eye view of the left-hand side of Eq.~(\ref{thirteen}) with
$\nu=0$ (measured in a unit of $g^3$) on a $20\times 10$ lattice. This was
obtained by set~IV~(c) of Table~\ref{table:1}.}
\label{fig:2}
\end{figure}

It is instructive to plot the left-hand side of WT
identities~(\ref{thirteen})--(\ref{sixteen}). If supersymmetry is restored, the
left-hand side must vanish except at~$x=0$ and its periodic images; recall
Eq.~(\ref{seventeen}). In~Fig.~\ref{fig:2}, we showed the left-hand side of
Eq.~(\ref{thirteen}) obtained by set~IV~(c) of Table~\ref{table:1}. Here, we
used the symmetric difference,
$\partial_\mu^{(\text{s})}f(x)\equiv\{f(x+a\hat\mu)-f(x-a\hat\mu)\}/(2a)$,
as a lattice transcription of the total divergence, because we found that its
use considerably reduces the statistical error.
We see that the left-hand side is almost zero everywhere (the absolute value is
less than~$0.05$ in the central portion) and it is sharply peaked at the
origin~$x=0$ and its periodic images. This result strongly indicates that our
reasoning in Sec.~\ref{sec:2} is correct and, at the same time, that
supersymmetry is certainly restored in the continuum limit. Recall that
Eq.~(\ref{thirteen}) is associated with the fermionic symmetry~$Q^{(0)}$ that is
\emph{not\/} preserved in the present lattice model with finite lattice
spacings.

\begin{figure}[htbp]
\begin{center}
\includegraphics*[width=\textwidth,clip]{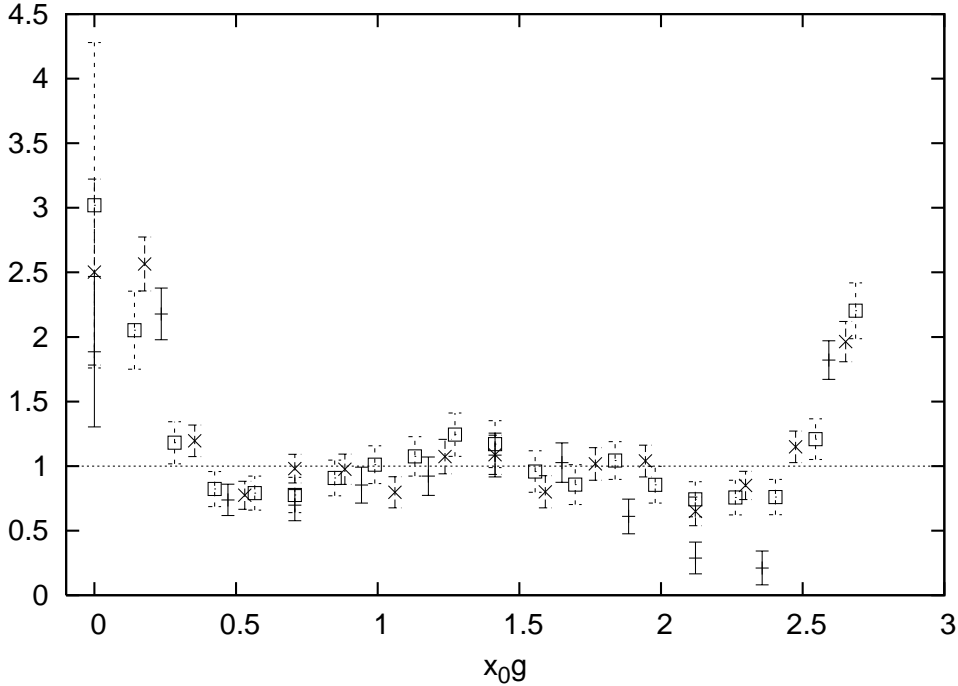}
\end{center}
\caption{Ratio~(\ref{twenty}) along the line~$x_1=L/2$. These were obtained
by set IV~(a) ($+$), set IV~(b)~($\times$) and IV~(c)~($\square$),
respectively.}
\label{fig:3}
\end{figure}

For systematic quantification of the supersymmetry restoration, we consider the
following four ratios:
\begin{gather}
   \frac{
   \partial_\mu^{(\text{s})}
   \left\langle\mathcal{J}_\mu^{(0)}(x)X_0^{(0)}(0)\right\rangle}
   {\left\langle Y^{(0)}(x)X_0^{(0)}(0)\right\rangle},
\label{twenty}\\
   \frac{
   \partial_\mu^{(\text{s})}
   \left\langle\mathcal{J}_\mu^{(1)}(x)X_0^{(1)}(0)\right\rangle}
   {\left\langle Y^{(1)}(x)X_0^{(1)}(0)\right\rangle},
\label{twentyone}\\
   \frac{
   \partial_\mu^{(\text{s})}
   \left\langle\widetilde{\mathcal{J}}_\mu(x)\widetilde X_0(0)\right\rangle}
   {\left\langle\widetilde Y(x)\widetilde X_0(0)\right\rangle},
\label{twentytwo}\\
   \frac{
   \partial_\mu^{(\text{s})}
   \left\langle\mathcal{J}_\mu(x)X_0(0)\right\rangle}
   {\left\langle Y(x)X_0(0)\right\rangle}.
\label{twentythree}
\end{gather}
According to PCSC relation (\ref{thirteen})--(\ref{sixteen}),
these ratios must become $\mu^2/g^2$ in the continuum limit, except at $x=0$
and its periodic images. We plotted ratio~(\ref{twenty}) for different lattice
spacings in Fig.~\ref{fig:3}. The statistical error in the ratio was estimated
by jackknife analysis. This plot is for $\mu^2/g^2=1.0$ and in the continuum
limit the points should lie on the dotted line. We see this expected
tendency, while a deviation near $x_0=0\equiv\beta$ can be understood as the
effect of approximate delta functions at~$x=0$ and periodic images elucidated
above. Also, again, we see that the lattice spacing $ag=0.2357$ is already
rather close to the continuum for $\mu^2/g^2=1.0$, because the WT identity is
fairly restored with this lattice spacing.

\begin{figure}[htbp]
\begin{center}
\includegraphics*[width=\textwidth,clip]{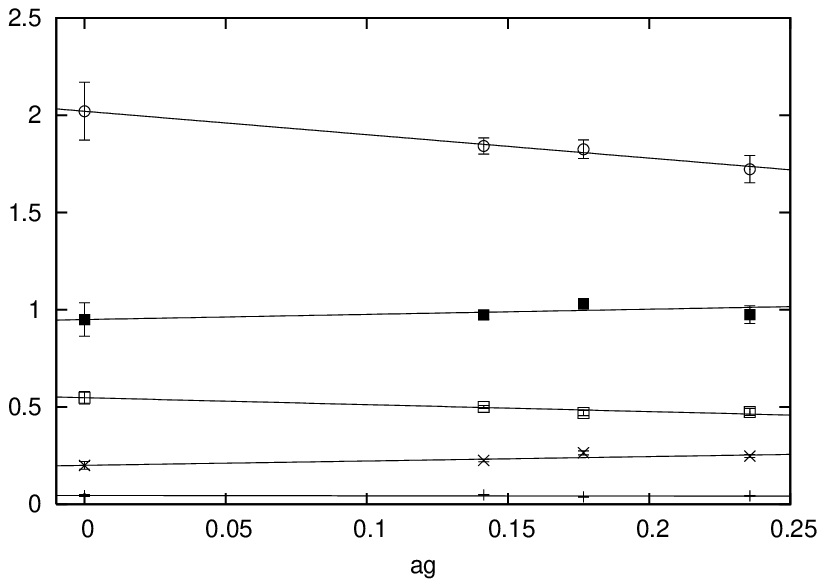}
\end{center}
\caption{The $\chi^2$-fitted values of ratio~(\ref{twenty}) as a function of
the lattice spacing~$ag$. Linear $\chi^2$-extrapolations to the
continuum~$a=0$ are also shown. Set I ($+$), set II~($\times$),
set III~($\square$), set IV~($\blacksquare$) and set V~($\bigcirc$).}
\label{fig:4}
\end{figure}

To quantify the restoration of supersymmetry, while eliminating the effect of
lattice artifacts existing around $x=0$ and its periodic images, we adopted
the following procedure. First, we take a cylindrical region~$C$ in~$\Lambda$,
$C\equiv\{x\in\Lambda\mid\beta/2-\Delta_1< x_0<\beta/2+\Delta_2,\,\,
0\leq x_1<L\}$. We then apply a $\chi^2$-fit by a constant to ratios
(\ref{twenty})--(\ref{twentythree}) obtained for $x\in C$. We change $\Delta_1$
and~$\Delta_2$ such that the number of points in the region~$C$ becomes maximum
while keeping the $\chi^2$ per one degree of freedom~(dof) reasonably small.
The results of this procedure are summarized in~Table~\ref{table:2} and
in~Fig.~\ref{fig:4}. In~Table~\ref{table:2}, columns indicated by ``used'' show
the number of points contained in the region~$C$ determined as above.
\begin{table}
\caption{The quality of the $\chi^2$-fit of ratios
(\ref{twenty})--(\ref{twentythree}) by a constant.}
\label{table:2}
\begin{center}
\begin{tabular}{cclclclcl}
\hline\hline
set label &
\multicolumn{2}{c}{Eq.~(\ref{twenty})} &
\multicolumn{2}{c}{Eq.~(\ref{twentyone})} &
\multicolumn{2}{c}{Eq.~(\ref{twentytwo})} &
\multicolumn{2}{c}{Eq.~(\ref{twentythree})} \\
\hline
 &
used & $\chi^2/\text{dof}$ &
used & $\chi^2/\text{dof}$ &
used & $\chi^2/\text{dof}$ &
used & $\chi^2/\text{dof}$ \\
\hline
I (a) & 12 & 1.533  & 12 & 0.7085 & 18 & 0.7947 & 12 & 1.062 \\
I (b) & 24 & 0.9475 & 48 & 0.9423 & 48 & 0.6633 & 24 & 0.5073\\
I (c) & 20 & 1.639  & 20 & 1.158  & 70 & 1.413  & 20 & 1.486 \\
\hline
II (a) & 18 & 0.9632 & 18 & 0.4774 & 18 & 0.7021 & 18 & 0.8510\\
II (b) & 16 & 0.8439 & 24 & 0.9131 & 16 & 0.6593 & 16 & 0.8143\\
II (c) & 50 & 0.9912 & 50 & 1.047  & 60 & 0.9219 & 50 & 0.9078\\
\hline
III (a) & 12 & 0.5984 & 18 & 0.6636 & 18 & 0.9439 & 18 & 0.5896\\
III (b) & 16 & 1.676  & 32 & 1.347  & 16 & 1.441  & 40 & 1.416 \\
III (c) & 80 & 0.9306 & 80 & 0.9628 & 70 & 0.9690 & 80 & 0.7924\\
\hline
IV (a) & 12 & 0.9697 & 18 & 0.8239 & 12 & 1.053  & 12 & 1.043 \\
IV (b) & 24 & 0.9034 & 32 & 0.9519 & 24 & 0.8340 & 24 & 0.7721\\
IV (c) & 50 & 0.9871 & 40 & 0.9419 & 50 & 0.9792 & 40 & 0.9535\\
\hline
V (a) & 18 & 0.6278 & 18 & 0.9550 & 12 & 0.6992 & 18 & 0.6746\\
V (b) & 32 & 0.9747 & 32 & 0.9105 & 16 & 0.7900 & 32 & 0.6894\\
V (c) & 60 & 1.469  & 80 & 1.326  & 70 & 1.423  & 70 & 1.346 \\
\hline\hline
\end{tabular}
\end{center}
\end{table}
In Fig.~\ref{fig:4}, the $\chi^2$-fitted values of ratio~(\ref{twenty})
obtained by the above procedure are plotted. As shown in the figure, these
values were then used for a linear $\chi^2$-extrapolation to the continuum.

\begin{figure}[htbp]
\begin{center}
\includegraphics*[width=\textwidth,clip]{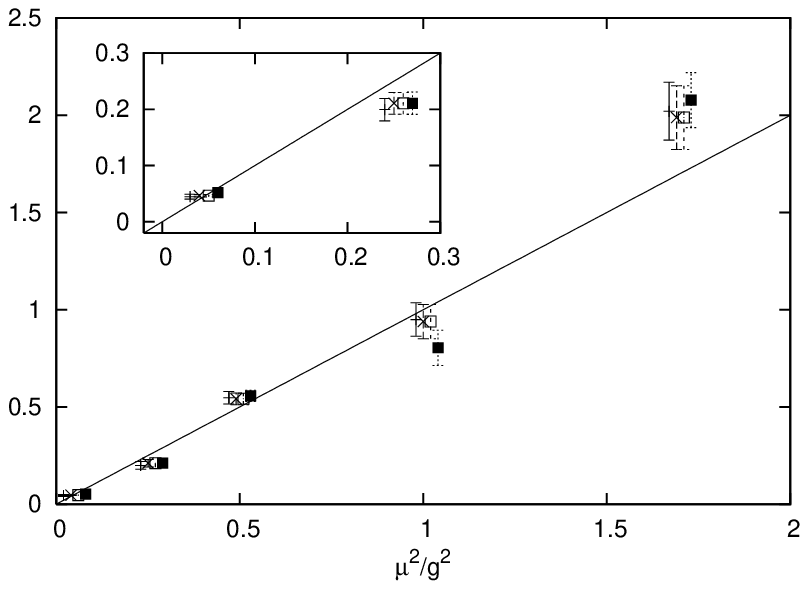}
\end{center}
\caption{The continuum limit of $\chi^2$-fitted values of ratios
(\ref{twenty})--(\ref{twentythree}) as a function of~$\mu^2/g^2$.
Eq.~(\ref{twenty})~($+$), Eq.~(\ref{twentyone})~($\times$),
Eq.~(\ref{twentytwo})~($\square$) and Eq.~(\ref{twentythree})~($\blacksquare$).
The straight line is a prediction of the PCSC relation. The plot
for~$\mu^2/g^2\leq0.3$ is magnified in the small window.}
\label{fig:5}
\end{figure}

Figure~\ref{fig:5} is the main result of this paper.
The continuum limit of ratios (\ref{twenty})--(\ref{twentythree}) obtained by
the above procedure is plotted as a function of the parameter~$\mu^2/g^2$. The
result is consistent with the straight line, that is a prediction of the
supersymmetric WT identities (the PCSC relation).

\begin{figure}[htbp]
\begin{center}
\includegraphics*[width=\textwidth,clip]{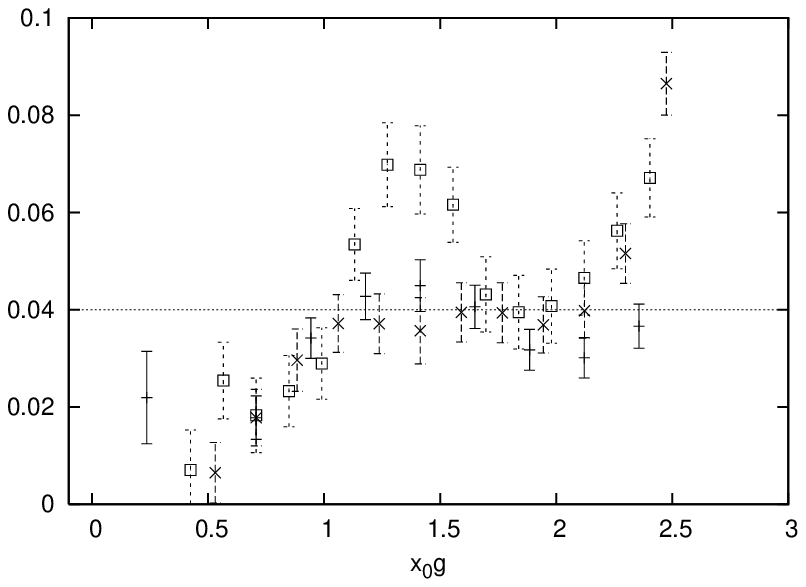}
\end{center}
\caption{Ratio~(\ref{twenty}) along the line~$x_1=L/2$. These were obtained
by set I~(a) ($+$), set I~(b)~($\times$) and I~(c)~($\square$),
respectively.}
\label{fig:6}
\end{figure}

Some care should be paid for the interpretation of Fig.~\ref{fig:5}; the plot
shows only statistical errors. There might be rather large systematic errors
associated with the fitting procedure especially for the smallest mass
$\mu^2/g^2=0.04$ case (set~I). Figure~\ref{fig:6} shows ratio~(\ref{twenty})
for set~I and we see that a flat region is narrower compared
with~Fig.~\ref{fig:3}, that is for~$\mu^2/g^2=1.0$.
In fact, as Table~\ref{table:2} shows, the numbers of points we used in the fit
are rather fewer for set~I than those for other sets. From Fig.~\ref{fig:6},
however, the systematic errors for set~I would be at most~$0.05$, thus the
result is still consistent with the PCSC relation even if this systematic error
is taken into account. For other values of the mass parameters~$\mu^2/g^2$,
the behavior of ratios is more or less similar to that in~Fig.~\ref{fig:3} and
the numbers of points we used in the fit appear sufficient; so we do not expect
large systematic errors.

Therefore, from the agreement with the theoretical expectation
in~Fig.~\ref{fig:5}, we infer that supersymmetry, that is broken in the present
lattice regularization with finite lattice spacings, is certainly restored in
the continuum limit.\footnote{As a simple check for that nontrivial loop
corrections are really important in our results, we repeated above analyses in
the quenched approximation. We confirmed that in fact the results significantly
differ from those obtained above and the expected supersymmetric WT identities
are not restored.}

The antiperiodic boundary condition explicitly breaks supersymmetry while the
periodic one preserves supersymmetry, so it is certainly of interest to carry
out simulations (for the energy spectrum, for example) with the latter boundary
condition. A typical example of correlation functions with the periodic
boundary condition is depicted in Fig.~\ref{fig:7}.
\begin{figure}[htbp]
\begin{center}
\includegraphics*[width=\textwidth,clip]{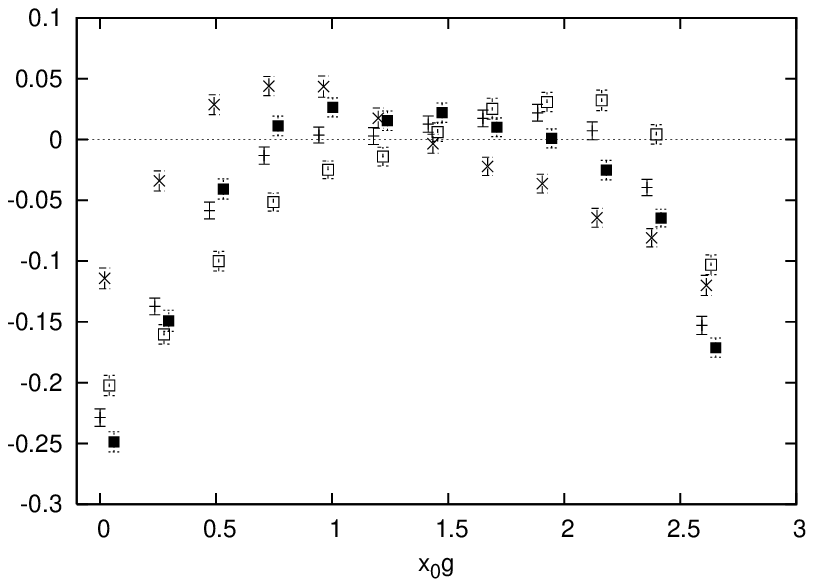}
\end{center}
\caption{Correlation functions~(\ref{eleven}) with $\nu=0$ along the
line~$x_1=L/2$. The lattice size is $12\times 6$ and~$ag=0.2357$.
$\mu^2/g^2=1.0$. The boundary condition is periodic. The number of
configurations is~800.
$\langle\mathcal{J}_0^{(0)}(x)X_0^{(0)}(0)\rangle/g^2$~($+$),
$\langle\mathcal{J}_0^{(1)}(x)X_0^{(1)}(0)\rangle/g^2$~($\times$),
$\langle\widetilde{\mathcal{J}}_0(x)\widetilde X_0(0)\rangle/g^2$~($\square$),
$\langle\mathcal{J}_0(x)X_0(0)\rangle/g^2$~($\blacksquare$).}
\label{fig:7}
\end{figure}
The degeneracy of four correlation functions is not quite realized. This
indicates that the spacing $ag=0.2357$ is not yet so close to the continuum for
$\mu^2/g^2=1.0$ when we use the periodic boundary condition. In addition to
this, we found that signals with the periodic boundary condition are generally
rather noisy and, when translated to the ratio like Fig.~\ref{fig:3}, the
statistical errors are too large for a reliable fit. For these reasons, we
postpone a detailed study of cases with the periodic boundary condition to a
future work.

\section{Subtlety in the supersymmetric $\mu^2=0$ case}
\label{sec:4}
In the above analyses, we have introduced the scalar mass term~(\ref{nineteen})
that generally suppresses the amplitude of scalar fields. One is of course
interested in the original two-dimensional $\mathcal{N}=(2,2)$ SYM that does
not contain such a supersymmetry breaking term. We now explain why we had to
introduce the scalar mass term in our simulation.

\begin{figure}[htbp]
\begin{center}
\includegraphics*[width=\textwidth,clip]{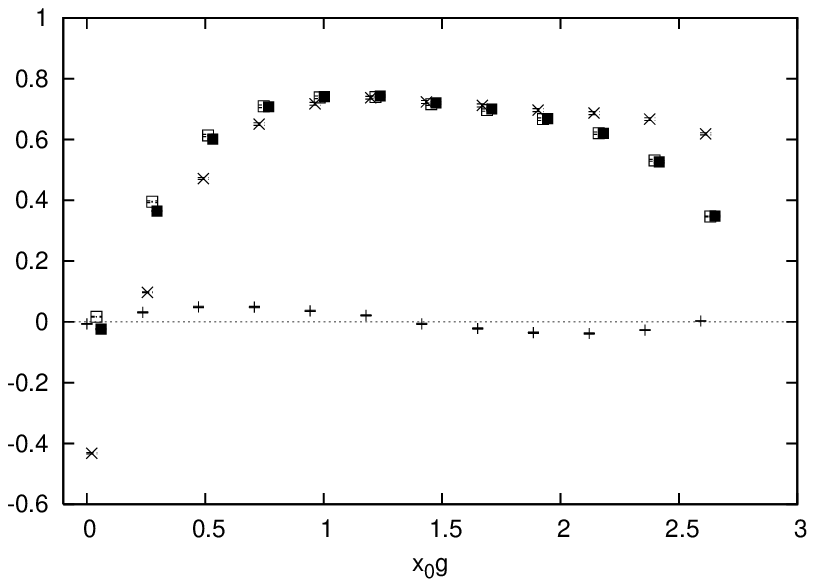}
\end{center}
\caption{Correlation functions~(\ref{eleven}) with $\mu=\nu=0$ along the
line~$x_1=L/2$. $\mu^2/g^2=0$. The lattice size is $12\times6$ and $ag=0.2357$.
The boundary condition is antiperiodic. The number of configurations is~800.
$\langle\mathcal{J}_0^{(0)}(x)X_0^{(0)}(0)\rangle/g^2$~($+$),
$\langle\mathcal{J}_0^{(1)}(x)X_0^{(1)}(0)\rangle/g^2$~($\times$),
$\langle\widetilde{\mathcal{J}}_0(x)\widetilde X_0(0)\rangle/g^2$~($\square$),
$\langle\mathcal{J}_0(x)X_0(0)\rangle/g^2$~($\blacksquare$).}
\label{fig:8}
\end{figure}

\begin{figure}[htbp]
\begin{center}
\includegraphics*[width=\textwidth,clip]{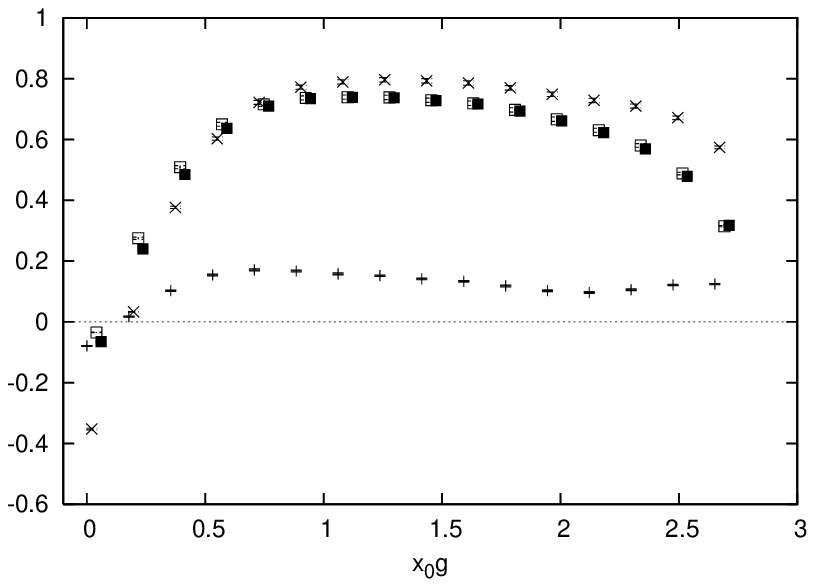}
\end{center}
\caption{Correlation functions~(\ref{eleven}) with $\mu=\nu=0$ along the
line~$x_1=L/2$. $\mu^2/g^2=0$. The lattice size is $16\times8$ and $ag=0.1768$.
The boundary condition is antiperiodic. The number of configurations is~800.
$\langle\mathcal{J}_0^{(0)}(x)X_0^{(0)}(0)\rangle/g^2$~($+$),
$\langle\mathcal{J}_0^{(1)}(x)X_0^{(1)}(0)\rangle/g^2$~($\times$),
$\langle\widetilde{\mathcal{J}}_0(x)\widetilde X_0(0)\rangle/g^2$~($\square$),
$\langle\mathcal{J}_0(x)X_0(0)\rangle/g^2$~($\blacksquare$).}
\label{fig:9}
\end{figure}

Figure~\ref{fig:8} shows correlation functions~(\ref{eleven}) without the
scalar mass term.
The lattice spacing is $ag=0.2357$. One sees that the degeneracy among four
correlation functions is enormously violated and thus it appears that we are
quite far from the continuum theory. Even if we decrease the spacing to
$ag=0.1768$ (Fig.~\ref{fig:9}), the crude feature does not change and the
degeneracy is not restored.
This behavior prompts us to draw a conclusion that the degeneracy is not
restored even in the continuum limit.

Moreover, from a closer look at Figs.~\ref{fig:8} and~\ref{fig:9}, it appears
that the non-degeneracy cannot be understood as a breaking of a certain
symmetry in the continuum theory. In these figures, only
$\langle\mathcal{J}_0^{(0)}(x)X_0^{(0)}(0)\rangle$ (indicated by~$+$) is quite
different from others while other three are almost degenerated. As noted
around~Eq.~(\ref{eleven}), these four correlation functions in the continuum
theory are related each other by certain global symmetries.\footnote{Note that
the boundary condition does not affect these global bosonic symmetries.} For
example, $\langle\mathcal{J}_0(x)X_0(0)\rangle$ and
$\langle\widetilde{\mathcal{J}}_0(x)\widetilde X_0(0)\rangle$ are related by
transformation~(\ref{seven}) with $\alpha=\pi/2$. The fact that these two
functions are quite degenerated in the figures suggests that this global
discrete symmetry is well restored with the present lattice spacings.
However, $\langle\mathcal{J}_0^{(0)}(x)X_0^{(0)}(0)\rangle$ and
$\langle\mathcal{J}_0^{(1)}(x)X_0^{(1)}(0)\rangle$ are not degenerated at all,
although these two are also related by this symmetry in the continuum theory.
Something quite strange
seems happening.

Another observation we made is that if we modify a term in the lattice Dirac
propagator that originates from the Yukawa interaction, by an $O(a)$ amount (in
the process of measurement), the effect of the modification is tremendous and
four correlation functions become almost degenerated; the effect appears to be
$O(1)$.

We suspect that the above strange behavior in the absence of the scalar
mass term is caused by \emph{very large\/} expectation value of scalar fields
along flat directions---continuous set of minima of the classical scalar
potential. Here, by ``very large'', we mean the lattice cutoff scale, $O(1/a)$.
In fact, as Fig.~3 of Ref.~\cite{Kanamori:2008vi} shows, the (gauge invariant)
amplitude of scalar fields can be as large as $\sim 40/a$, with the
antiperiodic boundary condition.\footnote{On the other hand, we numerically
observed that the amplitude does not grow so much ($\lesssim1/a$) with the
\emph{periodic\/} boundary condition. This seems strange at first glance
because the periodic boundary condition does not break supersymmetry and one
may expect that the flat directions are not lifted upon radiative corrections
when supersymmetry is preserved. This phenomenon could be understood if we
consider an effective potential for scalar zero modes that is obtained by
integrating out all other modes. With the antiperiodic boundary condition,
there is no fermion zero mode and the degrees of freedom that integrated out
are balancing between bosons and fermions. On the other hand, with the periodic
boundary condition, fermionic zero modes are left unbalanced and they can
induce nontrivial corrections. See~Refs.~\cite{Aoki:1998vn,Catterall:2007fp}
for related observations.}

Such a very large expectation value could ruin perturbative power counting,
in which the operator $a\phi$, for example, is regarded as $O(a)$. That is, if
the expectation value of the scalar field~$\phi$ is as large as~$O(1/a)$, the
combination~$a\phi$ would have to be regarded as $O(1)$. This phenomenon
affects also the argument for the supersymmetry restoration. For example, the
operator $Q(a\tr\{\overline\phi\psi_\mu\})
=a\tr\{\eta\psi_\mu\}+a\tr\{\overline\phi iD_\mu\phi\}$ is invariant under
the gauge, $U(1)_A$ and~$Q$ transformations---manifest symmetries of the
present lattice formulation---and thus could radiatively be induced (in the
one-loop level). This is an irrelevant operator in the usual sense but this
could be ``marginal'' if the scalar field~$\overline\phi$ is regarded as
$O(1/a)$. This $O(1)$ operator, moreover, is not invariant under $Q^{(0)}$,
$Q^{(1)}$ and~$\widetilde Q$ transformations, thus invalidates the argument for
the supersymmetry restoration.\footnote{One might think that the above operator
is prohibited to appear in the effective action owing to the lattice rotation
and reflection symmetries. We should note, however, that the present lattice
formulation does not possess such symmetries.}

It is thus conceivable that the target theory, the two-dimensional
$\mathcal{N}=(2,2)$ $SU(2)$ SYM, is not realized in the continuum limit of the
present lattice formulation, unless we supplement the scalar mass term (or
something that suppresses the amplitude of scalar fields). This point is
further discussed in the next section. Also, in light of this observation, a
study on the dynamical supersymmetry breaking in the two-dimensional
$\mathcal{N}=(2,2)$ $SU(2)$ SYM in Refs.~\cite{Kanamori:2007yx,Kanamori:2007ye}
must be reconsidered~\cite{Kanamori}.

\section{Conclusion and discussion}
\label{sec:5}
In this paper, for the first time to our knowledge, the restoration of
supersymmetry in a lattice formulation of a supersymmetric gauge theory was
clearly observed. The PCSC relation in~Fig.~\ref{fig:5}, first and
foremost, can be taken as a solid basis for the lattice model
in~Ref.~\cite{Sugino:2004qd} to be used for evaluation of various quantities in
the two-dimensional $\mathcal{N}=(2,2)$ SYM with a supersymmetry breaking
scalar mass term. Also, our result for a two-dimensional model demonstrates
validity of general reasoning for the supersymmetry restoration on the basis of
lattice symmetries and power counting. It is quite conceivable that, therefore,
various lattice formulations of supersymmetric gauge theory that are based on
similar reasoning work as they are aimed at.

The above statement is for the lattice model in which there is no flat
direction in the classical potential. The lattice model for the original
$\mathcal{N}=(2,2)$ SYM without the scalar mass term~\cite{Sugino:2004qd}
possesses flat directions and, as we have argued in~Sec~\ref{sec:4}, it is
quite plausible that the target theory is not obtained by the continuum limit.
Thus, we close this paper by summarizing the situation concerning the
$\mathcal{N}=(2,2)$ SYM without supersymmetry breaking terms:

(1)~One may first introduce the scalar mass term and then take the
limit~$\mu^2\to0$. Physically, as discussed in Ref.~\cite{Cohen:2003xe}, if the
mass $\mu$ is sufficiently small compared with the inverse of the system
size, $1/L$, the effect of the breaking mass would be practically negligible
because one cannot observe the wavelength longer than the size of the
``universe''. This provides a possible way of defining the $\mathcal{N}=(2,2)$
SYM. This route of definition would be mandatory in lattice formulations in
Refs.~\cite{Cohen:2003xe,D'Adda:2005zk,Catterall:2007kn}, in which a
supersymmetry breaking scalar mass term must be introduced to stabilize the
lattice spacing. Our result in Fig~\ref{fig:5} suggests that this prescription
works in the formulation of Ref.~\cite{Sugino:2004qd}, because we have observed
the restoration of supersymmetry even for $\mu^2\lesssim1/L^2=0.5g^2$.

(2)~One may carry out Monte Carlo simulations without introducing the
supersymmetry breaking mass term. This is possible for example in the
formulations of Refs.~\cite{Sugino:2004qd,Suzuki:2005dx}. In this approach,
however, there will be a subtle problem we encountered in Sec.~\ref{sec:4}
that $O(a)$ lattice artifacts seem to be amplified to $O(1)$ by $O(1/a)$
expectation value of scalar fields along flat directions.\footnote{In the
large~$N_c$ limit, there is a possibility that the expectation value of scalar
fields along flat directions is suppressed and one can evade the problem
without breaking supersymmetry. See Ref.~\cite{Anagnostopoulos:2007fw}.} Our
observation suggests that it is generally quite difficult to realize
supersymmetric theories with flat directions as a continuum limit of a lattice
model, without suppressing the amplitude of scalar fields, because lattice
formulation generally cannot be free from $O(a)$ discretization errors.

A profound question is, however, whether above prescription~(1) really provides
a ``correct'' definition of the target two-dimensional theory or not; the
existence of the vacuum moduli is unlikely in two dimensions while the
prescription would enforce scalar fields to localize around the origin.
Unfortunately, we do not have the convincing answer to this question at
present.

We would like to thank Masanori Hanada, Hikaru Kawai, Martin L\"uscher, Hideo
Matsufuru and Fumihiko Sugino for enlightening discussions. Our numerical
results were obtained using the RIKEN Super Combined Cluster (RSCC). I.~K.\ is
supported by the Special Postdoctoral Researchers Program at RIKEN. The work of
H.~S.\ is supported in part by a Grant-in-Aid for Scientific Research,
18540305.

\appendix

\section{Derivation of supersymmetric WT identities
(\ref{thirteen})--(\ref{sixteen})}
\label{appendix:A}
In this appendix, supersymmetric WT
identities~(\ref{thirteen})--(\ref{sixteen}) are derived with emphasize on
their independence of boundary conditions. We start with the expectation value
of operator~(\ref{ten}):
\begin{equation}
   \left\langle f_\nu(0)\right\rangle=\frac{1}{\mathcal{Z}}
   \int d\mu\,e^{-S-S_{\text{mass}}}f_\nu(0),
\end{equation}
where $\mathcal{Z}$ is the partition function and $d\mu$ denotes a measure of
the functional integral. We then consider a certain infinitesimal variation
of integration variables~$\delta'$,
$A_M\to A_M+\delta'A_M$, $\Psi\to\Psi+\delta'\Psi$ and~$H\to H+\delta'H$. 
Since the functional integral itself is independent of any relabeling of
integration variables, we have the identity for \emph{any\/}
variation~$\delta'$,
\begin{equation}
   \left\langle-i\delta'\left(S+S_{\text{mass}}\right)f_\nu(0)\right\rangle
   =-i\left\langle\delta'f_\nu(0)\right\rangle,
\label{axtwo}
\end{equation}
\emph{provided that\/} the measure $d\mu$ is invariant under the
variation~$\delta'$.

Now, as a particular form of~$\delta'$, we take the super transformation with a
replacement $\epsilon\to\epsilon(x)$, where $\epsilon(x)$ is a Grassmann-odd
spinor function with a finite support that does not overlap with the boundary;
$\delta'A_M=i\epsilon^T(x)C\Gamma_M\Psi$,
$\delta'\Psi=\frac{i}{2}F_{MN}\Gamma_M\Gamma_N\epsilon(x)
+i\widetilde H\Gamma_5\epsilon(x)$
and $\delta'H=-i\epsilon^T(x)C\Gamma_5\Gamma_MD_M\Psi
+\epsilon^T(x)C\Gamma_0D_1\Psi-\epsilon^T(x)C\Gamma_1D_0\Psi$. Under this
variation
\begin{align}
   &\delta'\left(S+S_{\text{mass}}\right)=i\int d^2x\,
   \left\{-\partial_\mu\epsilon^T(x)s_\mu(x)-\frac{\mu^2}{g^2}\epsilon^T(x)f(x)
   \right\}
\nonumber\\
   &\qquad+\frac{i}{g^2}\int d^2x\,\partial_\mu
   \tr\left\{
   \epsilon^T(x)C
   \left(
   -\frac{1}{2}\Gamma_A\Gamma_B\Gamma_\mu F_{AB}
   +2\Gamma_AF_{\mu A}
   +\Gamma_\mu\Gamma_5\widetilde H\right)\Psi
   \right\},
\end{align}
where combinations $s_\mu$ and $f$ are given by Eqs.~(\ref{nine})
and~(\ref{twelve}), respectively. In this expression, the second line
identically vanishes for any boundary condition because $\epsilon(x)=0$ at the
boundary. Similarly, in the first line, we may perform integration by parts
neglecting boundary terms again because $\epsilon(x)=0$ at the boundary.
Finally, setting $\epsilon(x)$ to be the delta function (times a Grassmann-odd
constant spinor), we have Eqs.~(\ref{thirteen})--(\ref{sixteen}) as particular
cases of~Eq.~(\ref{axtwo}). The assumption that the measure~$d\mu$ is invariant
under~$\delta'$ corresponds to, in the present formal treatment, the assumption
that regularization preserves supersymmetry.

\section{Multi time step acceleration in our simulation}
\label{appendix:B}
Let $S_{\text{B}}$ be the action consists only of gauge and scalar fields and
$S_{\text{PF}}$ the action of pseudofermions. $S_{\text{PF}}$ is bi-linear in
pseudofermions. In the molecular dynamics, if there exists a definite hierarchy
between force originates from~$S_{\text{B}}$ and that from~$S_{\text{PF}}$, and if
the latter is smaller than the former, one may reduce computational cost for
the latter (it is expensive requiring inversion of a Dirac operator) by using
different time steps for each of force. This is the multi time step
acceleration~\cite{Sexton:1992nu}. In this scheme, the time-evolution operator 
in one trajectory $\Delta\tau=n\epsilon$ is symbolically written as
\begin{equation}
   T(S_{\text{PF}},\epsilon/2)\left\{
   \left[T(S_{\text{B}},\epsilon/k)\right]^k
   T(S_{\text{PF}},\epsilon)
   \right\}^{n-1}
   \left[T(S_{\text{B}},\epsilon/k)\right]^k
   T(S_{\text{PF}},\epsilon/2),
\label{bxone}
\end{equation}
where $T(S,\epsilon)$ denotes the time-evolution operator with respect to the
action~$S$ in the time step~$\epsilon$. That is, one uses a $k$-times larger
time step for force from~$S_{\text{PF}}$.

\begin{figure}[htbp]
\begin{center}
\includegraphics*[width=\textwidth,clip]{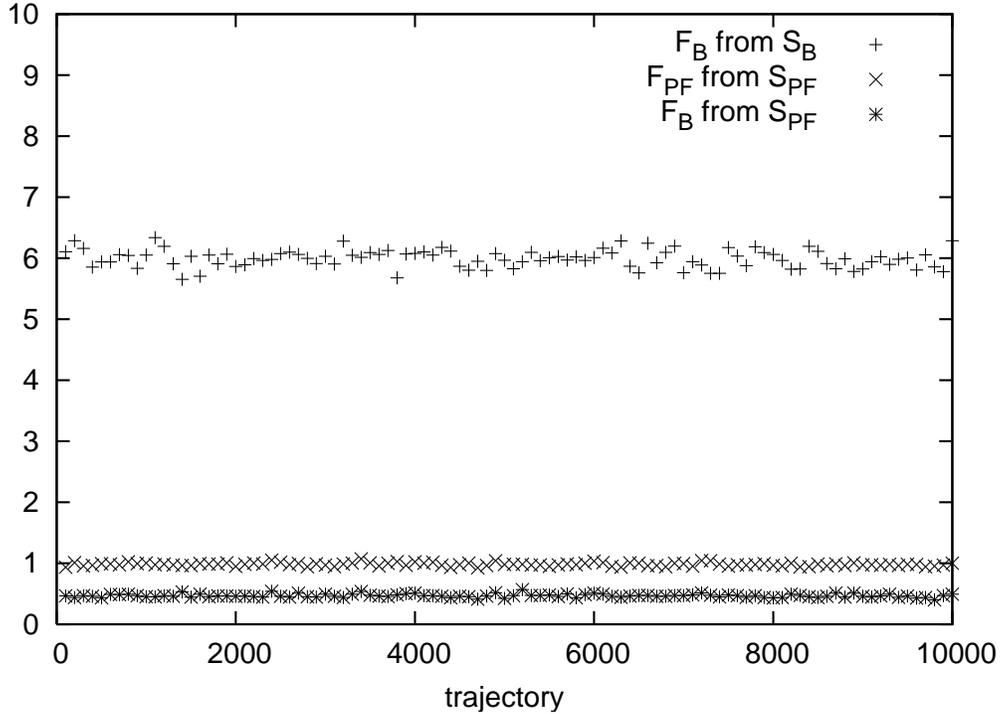}
\end{center}
\caption{Force for gauge and scalar fields ($F_{\text{B}}$) and that for
pseudofermions ($F_{\text{PF}}$), for a $12\times6$ lattice with~$ag=0.2357$
and~$\mu^2/g^2=0.25$. The values are averaged over each trajectory and plotted
every 100 trajectories. Force from~$S_{\text{B}}$ is about 6~times larger than
that from $S_{\text{PF}}$.}
\label{fig:10}
\end{figure}

In our Monte Carlo simulation, as shown in~Fig.~\ref{fig:10}, we found that
typically force from~$S_{\text{PF}}$ is several times smaller than that
from~$S_{\text{B}}$ and the scheme is in fact very efficient.
In the example in the figure, our choice was $k=3$ and $n=6$
in~Eq.~(\ref{bxone}) (we set $\Delta\tau=n\epsilon=0.5$). Since the variation
of the force is large (typically the same order of magnitude as the average
itself), we chose a smaller value of~$k$ than naively expected from the figure.
As a general tendency, when a lattice spacing becomes smaller, force from $S_B$
becomes larger and, correspondingly, we could use larger $k$. In fact, we used
$k=5$, $n=6$ for~$ag=0.1768$, and $k=8$, $n=6$ for $ag=0.1414$.

\end{document}